\documentclass[final,3p,times,twocolumn]{elsarticle}

\usepackage{pstricks} 
\usepackage{epstopdf}
\usepackage{subfigure}  

\usepackage{graphics}
\usepackage{graphicx}
\usepackage{epsfig}

\usepackage{amsmath}
\usepackage{lineno}

\usepackage{color,soul}    

\usepackage{amssymb}

\begin{document}

\begin{frontmatter}
\title{Towards accelerating Smoothed Particle Hydrodynamics simulations \\for free-surface flows on multi-GPU clusters}
\author[manchester]{Daniel~Valdez-Balderas}
\ead{Daniel.ValdezBalderas@manchester.ac.uk}
\author[vigo]{Jos\'e~M.~Dom\'{\i}nguez}
\author[manchester]{Benedict~D.~Rogers\corref{cor1}}
\ead{benedict.rogers@manchester.ac.uk}
\author[vigo]{Alejandro~J.C.~Crespo}
\cortext[cor1]{Corresponding author}
\address[manchester]{Modelling and Simulation Centre (MaSC), School of Mechanical, Aeroespace \& Civil Engineering, University of Manchester, Manchester, M13 9PL, UK}
\address[vigo]{Environmental Physics Laboratory, Universidad de Vigo, Ourense, Spain}
\begin{abstract}

Starting from the single graphics processing unit (GPU) version of the
Smoothed Particle Hydrodynamics (SPH) code DualSPHysics, a multi-GPU
SPH program is developed for free-surface flows. The approach is based
on a spatial decomposition technique, whereby different portions (sub-domains) of
the physical system under study are assigned to different
GPUs. 
%
%
Communication between devices is achieved with the use of Message
Passing Interface (MPI) application programming interface (API) routines.
The use of the sorting algorithm radix sort for inter-GPU particle
migration and sub-domain ``halo'' building (which enables interaction
between SPH particles of different sub-domains) is described
in detail.
%
%
With the resulting scheme it is possible,
on the one hand, to carry out simulations that could also be performed on a
single GPU, but they can now be performed even faster than on one of
these devices alone. On the other hand, accelerated simulations can be
performed with up to 32 million particles on the current architecture,
which is beyond the limitations of a single GPU due to memory
constraints. A study of weak and strong scaling behaviour, speedups
and efficiency of the resulting program is presented including an investigation
to elucidate the computational bottlenecks. Last, possibilities for
reduction of the effects of overhead on computational efficiency in
future versions of our scheme are discussed.


\end{abstract}

\end{frontmatter}

\section{Introduction}
\label{sec:introduction}

The applicability of particle-based simulations is typically limited
by two different but related computational constraints: simulation
time and system size. That is, to obtain physically meaningful
information from a simulation, one must be able to simulate a
large-enough system for long-enough times.  In the particular case of
the Smoothed Particle Hydrodynamics (SPH) method, certain types of
applications, for example the study of coastal processes and flooding
hydrodynamics, have been limited until now by the maximum number of
particles in order to perform simulations within reasonable times.

To overcome these limitations, various types of
acceleration techniques have been employed, which can
  be grouped into three main categories based on the type of hardware
used. On the one hand there are the traditional High Performance
Computing (HPC) techniques which involve the use of hundreds or
thousands of computing nodes, each hosting one or more Central
Processing Units (CPUs) containing one or more computing cores. 
Those nodes are interconnected via a computer
networking technology (e.g., Ethernet, Infiniband, etc.), and
programmed with the help of protocols like the Message Passing
Interface (MPI). For SPH, examples of this type of approach include
the work of Maruzewski \emph{et al.}
\cite{LMH-ARTICLE-2009-002}, who carried out SPH simulations with up
to 124 million particles on as many as 1024 cores on the IBM Blue
Gene/L supercomputer. Another recent example in this field is that of
Ferrari \emph{et al.} \cite{Ferrari20091203}, who reported
calculations using up to 2 million particles on a few hundred
CPUs. The drawback of this type of approach comes from the fact that,
for SPH, an enormous number of cores is needed, which require 
considerable investment, including the purchase, maintenance, and
power supply requirements of this type of equipment.

\begin{figure*}
  \centerline{\subfigure[]{\includegraphics[width=3.8in]{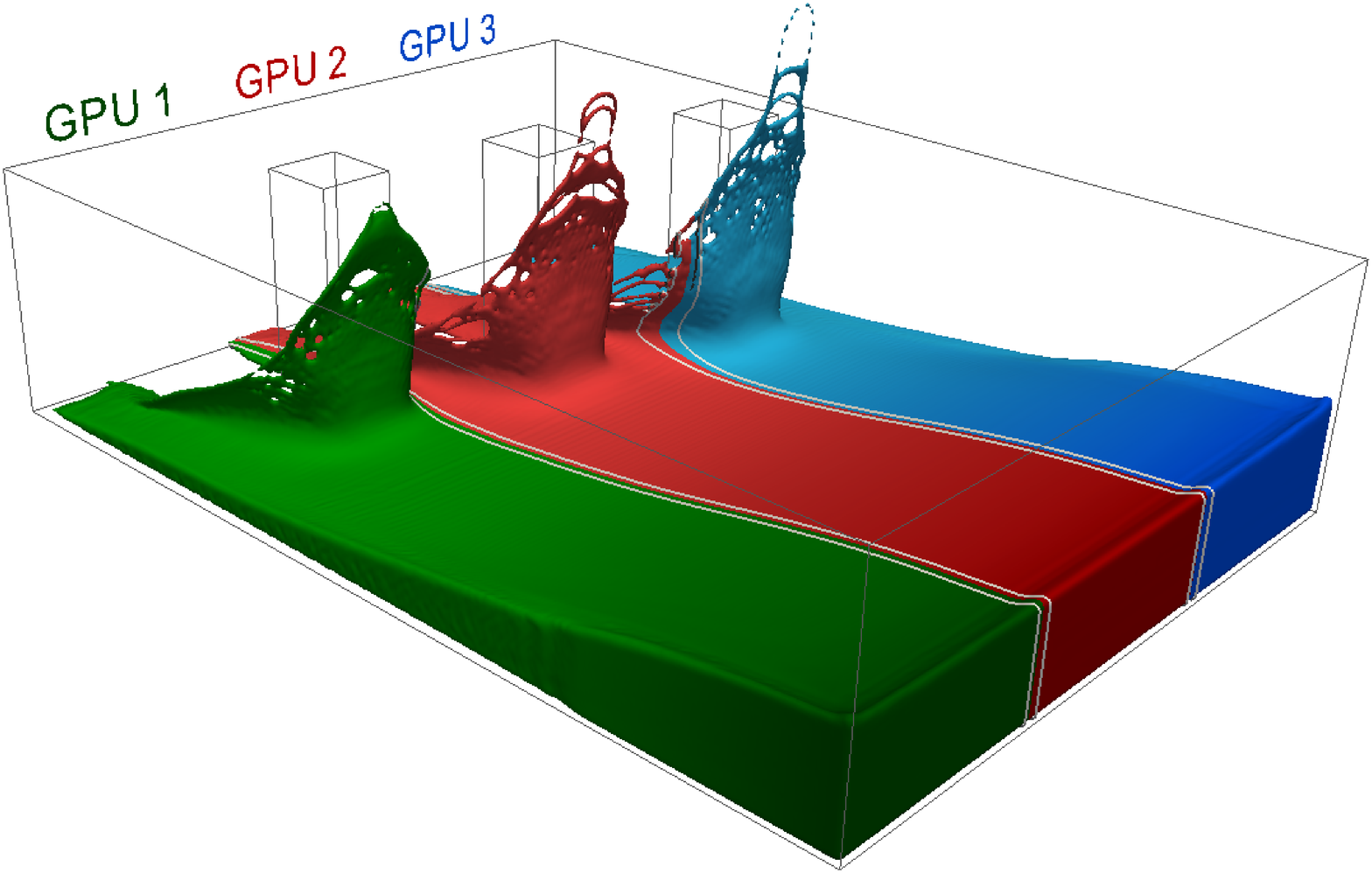}
      \label{fig_first_case}}
    \hfil
    \subfigure[]{\includegraphics[width=2.2in]{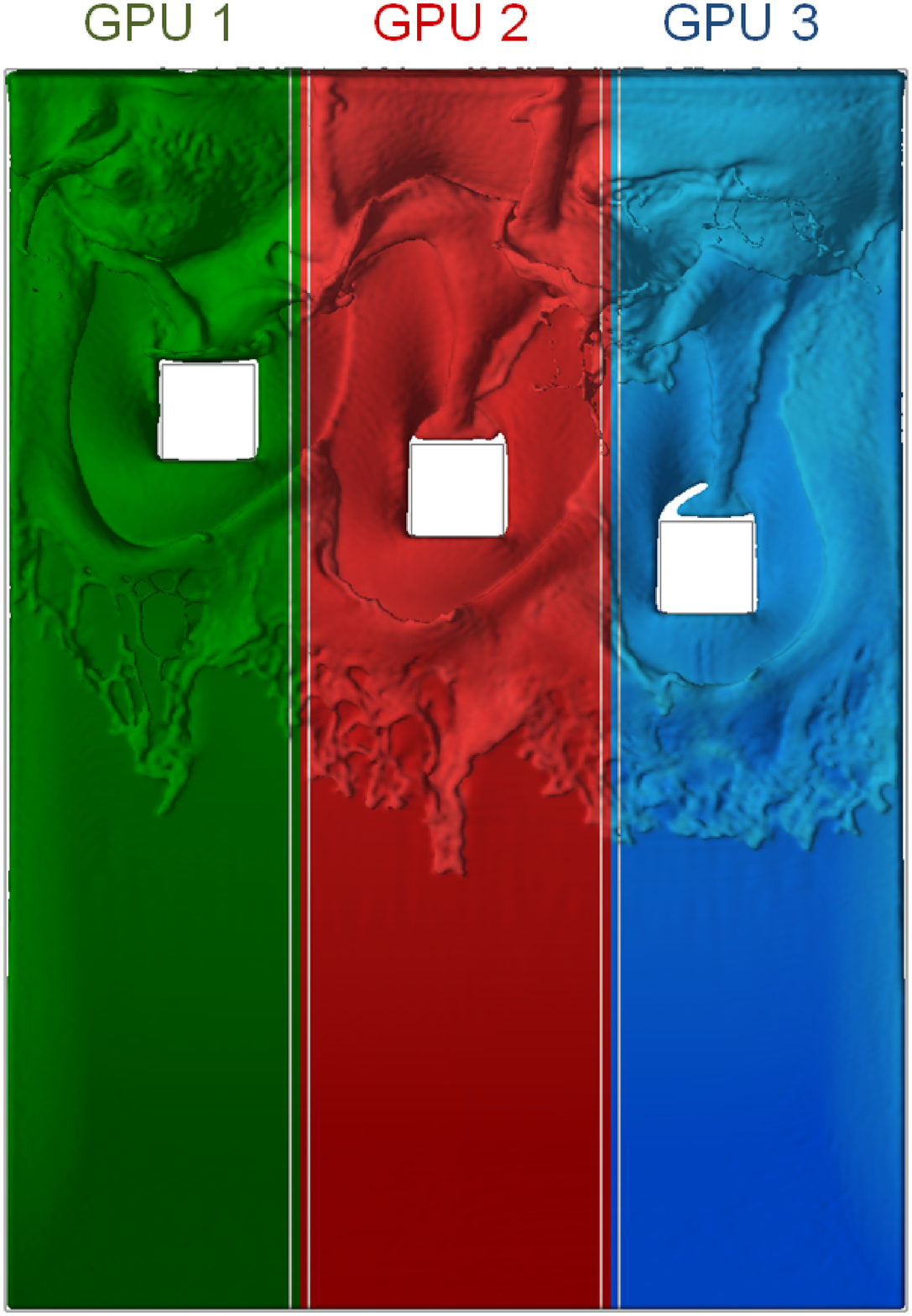}
      \label{fig_second_case}}}
  \caption{Snapshots of a multi-GPU SPH simulation using three GPUs,
    for a dam break with three obstacles. Portions of fluid in different
    sub-domains are displayed with a different color, and white lines
    near the sub-domain boundaries correspond to the halos.}
  \label{fig:ima_DANIEL0050}
\end{figure*}

A second type of acceleration approach is that involving the use of
Field-Programmable Gate Arrays (FPGAs). For example, Spurzem \emph{et
  al.} \cite{spurzemberczik2009} carried out SPH and gravitational
N-body simulations using this type of technology for astrophysics
problems, finding that it is useful for acceleration of complex
sequences of operations like those used in SPH. Since the main use
of FPGAs in the literature is in the field of astrophysics (where
long-range forces and variable smoothing lengths are typically
employed), the use of FPGAs for free-surface flow applications (with
short-range forces and fixed smoothing lenghts) remains a relatively
unexplored field.

The third type of acceleration technique used in SPH simulations, and
the one on which this article focuses, involves the use of a type of
hardware different from the CPU: the Graphics Processing Unit
(GPU). The development of GPU technology is driven by the computer
games industry but has recently been exploited for non-graphical
calculations, leading to the development of general purpose
GPUs (GPGPUs). GPU programming is a parallel approach because each of
these devices contains hundreds of computing cores, and multiple
threads of execution are launched simultaneously. The use of GPUs for
scientific computations has come to represent an exciting alternative
for the acceleration of scientific computing software. The release of
the Compute Unified Device Architecture (CUDA) and its software
development kit (SDK) by NVIDIA in 2007 has facilitated the
popularization of the use of these devices for general purposes, but
efforts in this direction existed even prior to that date. For
example, as early as 2004, Amada \emph{et al.}  \cite{amada2004}
carried out SPH simulations for real-time simulations of water. In
2007 Harada \emph{et al.}  \cite{harada2007sphongpus} reported SPH
simulations that ran an order of magnitude faster on GPUs than on
CPUs. More recently H\'{e}rault \emph{et al.}  \cite{herault2010},
reported one to two orders of magnitude speedups of the GPU when
compared to a single CPU.  Among the recent efforts for SPH
computations on GPUs, DualSPHysics \cite{website:dualsphysics} has
proven to be a versatile computational SPH code for both CPU and GPU
calculations. The GPU version of DualSPHysics maintains the stability
and accuracy of the CPU version, while providing 
significant speedups over the latter.
For a detailed description
of DualSPHysics, please refer to~\cite{crespo2011}.

Recently, in addition to performance, the energy efficiency of
different types of hardware is representing an increasingly important
factor when choosing hardware for accelerating 
simulations. This can be seen, for example, in the list of the fastest
supercomputers~\cite{top500}, which now include energy efficiency
along with performance specifications. In this realm, too, GPUs show
promise. For example, a recent article by McIntosh-Smith \emph{et al.}
~\cite{mcintoshsmith2011}, describing a methodology for energy
efficiency comparisons, reports that in a particular case study of
simulations of molecular mechanics-based docking applications, GPUs
delivered both better performance and higher energy efficiency than
multi-core CPUs.

However, GPU technology has its own constraints. In the particular
case of SPH, the memory restricts the maximum number of SPH particles
that can be efficiently simulated on a single device to approximately ten million or less, depending on the particular GPU and on the
precision of the data types used (e.g., single or double).  To go
beyond this limit, this work extends DualSPHysics by introducing a spatial
decomposition scheme with the use of the Message Passing Interface
(MPI) protocol\footnotemark to obtain a multi-GPU SPH application.
Thus, our approach combines the two types of parallelization
described above since it provides parallelization, first, at a coarse
level by decomposing the problem and assigning different volume
portions of the system to different GPUs, and second, at a fine level,
by assigning one GPU thread of execution to each SPH particle.  Our
resulting multi-GPU SPH scheme, illustrated in
  Fig.~\ref{fig:ima_DANIEL0050}, has the potential to bypass both the
system size and simulation time limitations which have to date
constrained the applicability of SPH in various engineering fields.

\footnotetext{This work started prior to the release
    of CUDA 4.0, which allows direct communications among multiple
    GPUs. See Section~\ref{sec:comptechmultip} for more details.}

The rest of this article is organized as follows: Section
\ref{sec:sphbasics} presents a brief overview of the main ideas behind
SPH. In Section \ref{sec:comptechsingle} the methodology used to
implement the single-GPU version \cite{crespo2011} (which is the starting point for the
present work) is summarized.  Section
\ref{sec:comptechmultip} describes our spatial decomposition, multi-GPU
algorithm, whereby the physical system is divided into sub-domains,
each of which is assigned to a different GPU. Section
\ref{sec:hardware} describes the hardware.  Section \ref{sec:results}
presents the main results of our simulations using a small number of
GPUs, including a study of weak and strong scaling, the bottlenecks of
the program, and our strategy to diminish the effect of those
bottlenecks and improve efficiency. Section \ref{sec:summary} concludes with a
summary and a description of ongoing and future work.

\section{Smoothed Particle Hydrodynamics}
\label{sec:sphbasics}

In order to better understand the specific challenges posed
  by a multi-GPU implementation of Smoothed Particle Hydrodynamics
  (SPH), a brief description of this method is now presented.  SPH is a
mesh-free, Lagrangian technique particularly well suited to study
problems that involve highly non-linear deformation of fluid surfaces
that occur in free-surface flows, such as wave breaking and marine
hydrodynamics~\cite{Vignjevic2010}. Developed originally for the study of astrophysical
phenomena, it now enjoys popularity in a variety of engineering
fields, such as civil engineering (e.g., the design of coastal
defenses), mechanical engineering (e.g., impact fracture in solid
mechanics studies), and metallurgy (e.g., mould filling).

The problem in SPH consists of determining the evolution in time of
the properties of a set of particles representing the fluid. In
engineering applications, particles have short range interactions with
their neighbours, and the dynamics are governed by a set of
simultaneous ordinary differential equations in time. In this section
the classical SPH formulation used in our simulations are described
(see ~\cite{GomezGesteira2010}).

\subsection{Governing equations}

The starting point for the derivation of the SPH equations is the set
of equations for the continnum description of dynamic fluid flow,
namely, the Navier-Stokes equations \cite{liuandliubook}:
%
\begin{equation}
 \frac{D\rho}{Dt}=-\rho \nabla \cdotp \vec v , \label{eq:navier1}
\end{equation}

\begin{equation}
 \frac{D\vec v}{Dt}=-\frac{1}{\rho}\nabla p + \vec g, \label{eq:navier2}
\end{equation}

\begin{equation}
 \frac{D e}{Dt}=-\frac{p}{\rho}\nabla \cdotp \vec v. \label{eq:navier3}
\end{equation}
Here, $\rho$ is the density, $t$ is time, $\vec v$ and $\vec
r$ represent velocity and position,
respectively, $\sigma$ is the total stress tensor, $e$ is the energy, $\vec g$ is the force due to gravity, and $D/Dt$
represents the material or total time derivative.

\subsection{SPH discretization of the Navier-Stokes equations}

Two key steps are used to discretize the Navier-Stokes equations:
first, the integral representation of a field function, or {\it
  kernel} approximation, and second, the {\it particle} approximation.
The value of a field function $f(\vec r)$ at point $\vec r$ can be
approximated by a weighted interpolation over the
  neighbouring volume in the following way:
\begin{equation}
  f(\vec r)\simeq \int_{\Omega} f(\vec r^{\prime})W(\vec r-\vec r^{\prime},h)d \vec r^{\prime},
  \label{eq:integralrep}
\end{equation}
where $W(\vec r-\vec r^{\prime},h)$ is a smoothing function (also called the smoothing kernel), $h$ is
the smoothing length (used to characterize the shape and range of 
$W$), and the integral is over all space $\Omega$.

\begin{figure*}
  \centerline{\subfigure[]{\includegraphics[width=2.2in]{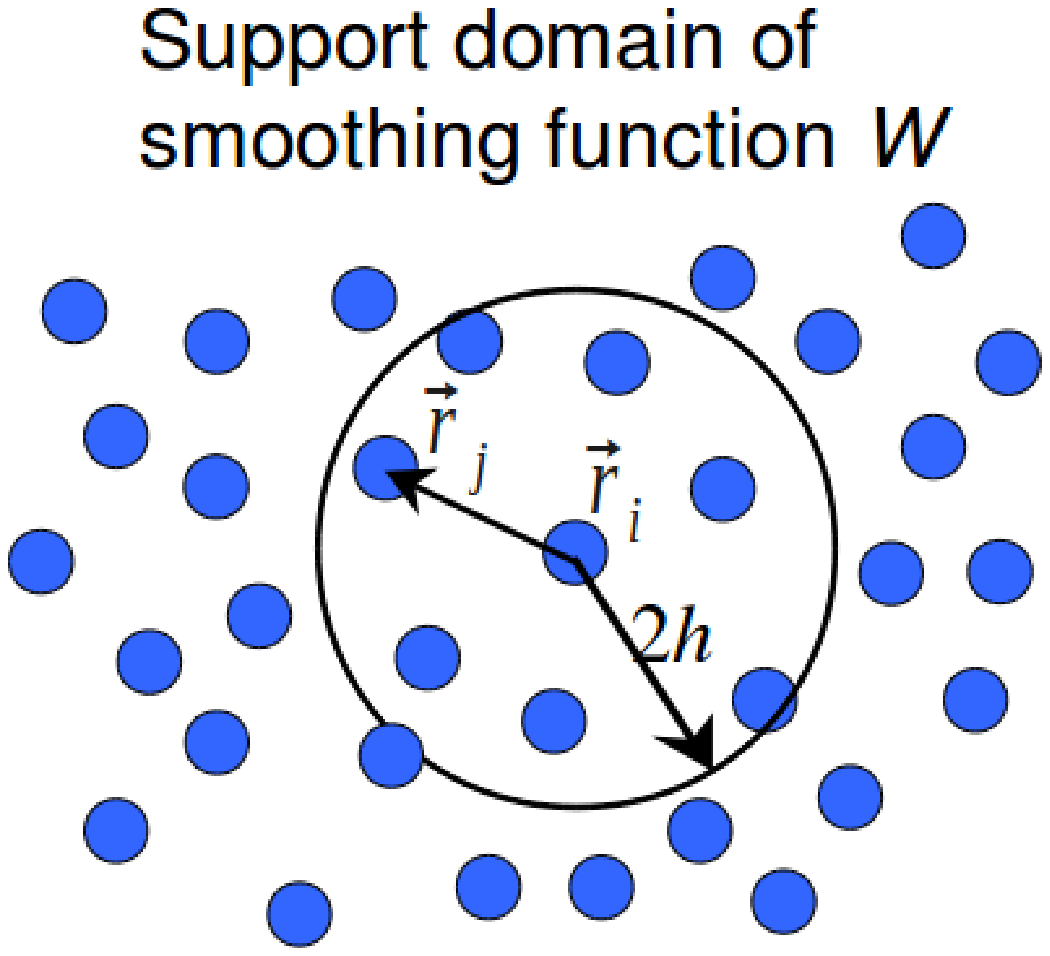}
      \label{fig:sphkernela}}
    \hfil
    \subfigure[]{\includegraphics[width=1.7in]{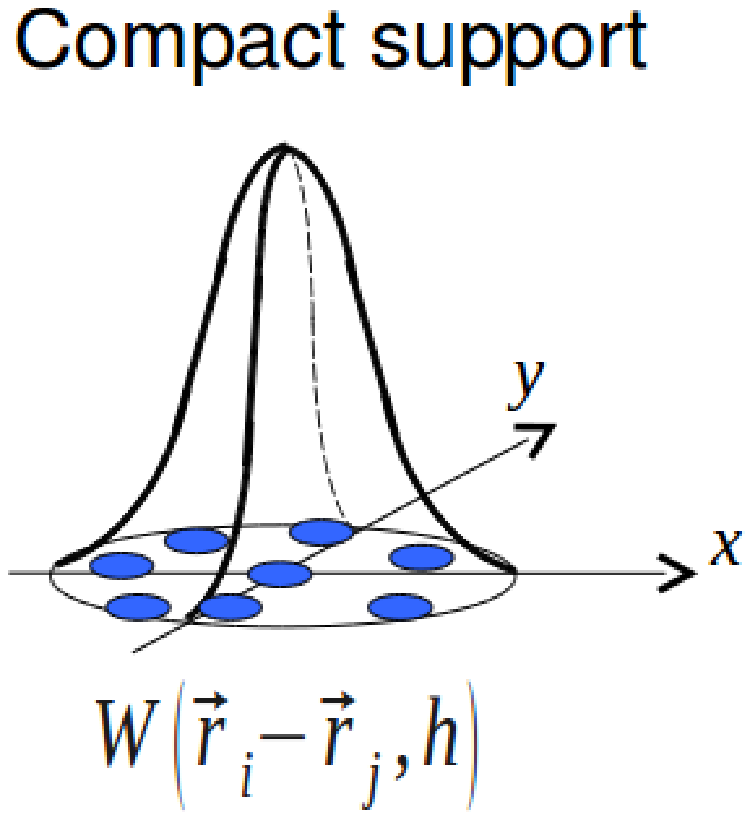}
      \label{fig:sphkernelb}}}
  \caption{Illustration of the smoothing function, $W(\vec r_i-\vec
    r_j, h)$, with a support domain of size $2h$. The blue circles are
    the particles that represent the fluid. Particle $i$ interact with
    particle $j$ only if $|\vec r_i-\vec r_j|<2h$, due to the compact
    support property of the smoothing function, as explained in the
    text.}
  \label{fig:sphkernel}
\end{figure*}

The smoothing function is typically chosen to have the following
properties: its integral over all space is unity; it
approaches the Dirac delta function as $h\rightarrow 0$; it has
compact support, i.e., its value is zero beyond its {\it support
  domain} $|\vec r-\vec r^{\prime}| > \kappa h$, where $\kappa$ is a
scaling factor used to set the coarseness of the discretization
of space (throughout this article
  $\kappa=2$); it monotonically decreases as $|\vec r-\vec
r^{\prime}|$ increases; and, last, it is symmetric
$W(\vec r-\vec r^{\prime},h)=W(|\vec r-\vec r^{\prime}|,h)$.  It can
be shown that the kernel approximation is accurate to
  order two in $h$, $\mathcal{O}(h^2)$\cite{monaghan1982}.  In this
article, all of the results are obtained using the cubic spline
smoothing function~\cite{liuandliubook} with a constant\footnotemark support domain of size
$2h$.

\footnotetext{Note that, in contrast with SPH
    simulations in astrophysics, e.g., Acreman \emph{et al.}
    \cite{acreman2010.1143A}, a variable smoothing length $h$ is not
    used here, since it is rarely required for simulations of
    free-surface flows. A variable smoothing length
    would increase the severity of diverging branches, load unbalance
    among threads, and irregular memory accesses, making an efficient
    GPU implementation of SPH a more challenging task than a fixed
    smoothing length implementation, like the one described in this
    article.}

The second step for the discretization of the Navier-Stokes equations is
the particle approximation, which consists simply of replacing the
continuous integral in Eq.~(\ref{eq:integralrep}) by a discrete sum,
and writing the differential of volume as a finite volume $\Delta V_j$
in terms of density and mass, obtaining
\begin{equation}
  f(\vec r_i)\simeq \sum_{j} f(\vec r_j)W(\vec r_i-\vec r_j,h)\frac{m_j}{\rho_j}
\end{equation}
where $m_j=\Delta V_j \rho_j$ is the mass of the SPH particle, and the
sum is performed over all neighboring particles in the support domain
of $W$.  

Figure~\ref{fig:sphkernel} illustrates some of the
  properties of the smoothing function $W(\vec r_i-\vec r_j,h)$. Part
  (a) of this figure shows particle $i$ interacting with all other
  particles, $j$, within its radius of influence, or support domain,
  whereas part (b) shows the compact support in operation, the smoothness of the
  function, as well as its monotonically decreasing behaviour for
  increasing $|\vec r_i-\vec r_j|$.

Following a similar argument, the particle approximation for
the spatial derivative of a function can be written in the following
form:
\begin{equation}
  \nabla f(\vec r_i)\simeq \sum_{j} [f(\vec r_j) - f(\vec r_i)] \nabla_iW(\vec r_i-\vec r_j,h)\frac{m_j}{\rho_j}
\end{equation}
Mathematically, therefore, the problem consists of performing
localised interpolations or summations around each computation point
where the properties of the fluid are evaluated.

In the present article, variations in the thermal properties of the
fluid are neglected, governed by
  Eqn.~(\ref{eq:navier3}), as the primary application of our approach
is free-surface flows where thermal properties do not play a
significant role.  Applying the SPH formulation to the governing
equations (\ref{eq:navier1}-\ref{eq:navier2}) leads to the following
set of equations~\cite{liuandliubook}:
\begin{eqnarray}
    \frac{d \rho_i(t)}{dt}    &  =  &   \sum_j m_j  \vec v_{ij} \cdotp \nabla_i W_{ij}\label{eq:sphdeneq}\\
    \frac{d \vec r_i(t)}{dt}  &  =  &   \vec v_i(t) \label{eq:sphposeq}\\
    \frac{d \vec v_i(t)}{dt}  &  =  &  -\sum_j m_j \left( \frac{p_i}{\rho_i} + \frac{p_j}{\rho_j} + \Pi_{ij}\right)\nabla_i W_{ij}+\vec g \label{eq:sphmomeq}
\end{eqnarray}
where $W_{ij} \equiv W(\vec r-\vec r_j,h)$, $\vec g$ is gravity, and
the artificial viscosity is given by $\Pi_{ij}=\frac{-\alpha \overline
  c_{ij} \nu_{ij}}{\rho_{ij}} \,\text{ if } \vec r_{ij}\cdotp \vec
v_{ij}<0$ and $\Pi_{ij}=0$ otherwise, where $\nu_{ij}=h\vec
v_{ij}\cdotp \vec r_{ij}/(\vec r_{ij}^2+\eta^2)$, $\vec r_{ij}\equiv
\vec r_i-\vec r_j$, $\vec v_{ij} \equiv \vec v_i-\vec v_j$, $
\overline c_{ij} \equiv (c_i+c_j)/2$, $\eta=0.1h$,
$\alpha$ is a parameter that can be related to the
  Reynolds number for the specific free-surface problem, and $p_i$ is
the pressure at $r_i$, which is governed by the equation of state $
p_i=B [ ( \rho_i/\rho_0 )^{\gamma}-1]$.  Here, $\gamma=7$,
$B=c_0^2\rho_0/\gamma$, $\rho_0=1000$ Kg/m$^3$, and
$c_{0}=\sqrt{(dp/d\rho)|_{\rho_0}}$.

\subsection{Boundary conditions}

In this article dynamic boundary conditions \cite{crespo2007} are used
to represent solid boundaries. In this method, boundary particles obey
the same dynamical equations as the fluid particles, but the
integration update for position and velocity is not
performed. Therefore, the density and pressure of the particle varies,
but the position and veolicty remain fixed (or change in an externally
imposed fashion in cases where solid boundaries are moving.)

\section{Single-GPU implementation}
\label{sec:comptechsingle}




To describe our multi-GPU implementation, 
it is useful to review the main strategy behind the single-GPU version, as
well as other general considerations that need to be taken into
account when performing SPH simulations. For a more detailed description of
the single-GPU version of DualSPHysics please refer to
\cite{website:dualsphysics} and \cite{crespo2011}.

Single-GPU DualSPHysics is a fully-on-GPU program, meaning that all
the calculations, to be described in detail below,
are performed on the GPU, and data resides on the GPU
memory at all times, and is copied to the host only when saving of the
simulation results is required. In this way the time consuming
transfers of data between CPU and GPU are minimized. This fully-on-GPU
strategy contrasts with other approaches which only partially port
computations to the GPU~\cite{ogerSPHERIC2010}.

Some, but not all, of the tasks involved in the GPU implementation of an
SPH simulation are easily parallelized. The difficulties mainly arise
from the Lagrangian nature of the method, where particles move in
space. The following basic strategies were followed to increase
performance: minimization of GPU-CPU data transfers; optimization of
memory usage to increase bandwidth; minimization of divergent warps;
optimization of memory access patterns; and maximization of occupancy to
hide latency.

An SPH simulation consists of solving the set of equations
(\ref{eq:sphdeneq})-(\ref{eq:sphmomeq}) at discrete points in time. To
achieve this, a series of iterations is performed:
\begin{enumerate}
\item Find the neighbours of each particle, that is, all other particles within its support domain of $2h$
\item Sum the pairwise interactions between each particle and its neighbours
\item Update the system with the use of an integrator
\end{enumerate}
We now describe the GPU implementation of those steps.

\subsection{Finding particle neighbours}

To determine the neighbours of each particle efficiently, the domain
space is divided into cubic cells of the size of the kernel support domain,
namely $2h$. Then a CUDA kernel is launched, with one thread per
particle, that assigns to each particle a label corresponding to the
cell to which it belongs. The next step consists of using the {\it
  radix sort} algorithm \cite{radixsortpaper2009} to order the array
holding the particle identification labels, or IDs, according to the
cell to which the particle belongs. Next all arrays containing data
information (position, velocity, etc.) are ordered according to the
newly ordered ID array. Last, an extra array is generated with the
index of the first particle of each cell that enables
neighbour identification during the computation of the particle interactions.


\begin{figure}
  \centering
  \includegraphics[height=6.0cm,angle=0]{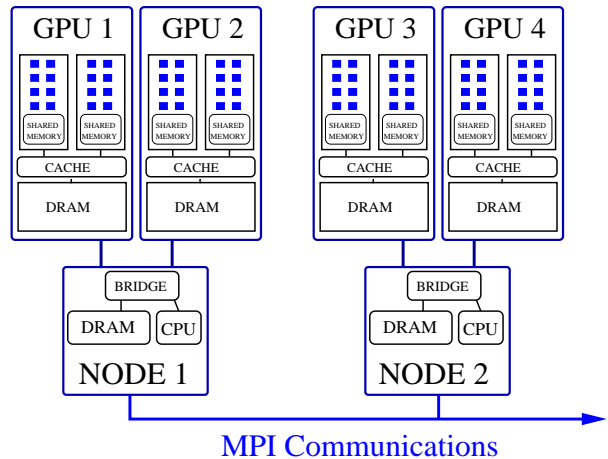}
  \caption{Illustration of a multi-GPU system with two nodes, each
    hosting two GPUs. More generally, a multi-GPU system 
    consists of one or more computing nodes, each hosting one or more
    GPUs, in addition to one or more CPUs. Different nodes are
    connected via a computer networking technology, and transfer of
    information is performed with MPI.}
  \label{fig:gpucpuconnected}
\end{figure}

\subsection{Computing particle interactions}

With the arrays re-ordered according to cells, and the index of the
first particle of each cell, a new CUDA kernel is launched. One CUDA
thread is assigned to each particle $i$ to search for its
neighbours. If particle $i$ is in cell $c$, the thread will look for
neighbours only in cell $c$ and in cells adjacent to $c$. This CUDA
thread will also compute the interaction between particle $i$ and its
neighbours, that is, the sum on the right-hand side of equations
(\ref{eq:sphdeneq})-(\ref{eq:sphmomeq}).

\subsection{System update}

Once the interactions are computed, and the right-hand side of equations
(\ref{eq:sphdeneq})-(\ref{eq:sphmomeq}) is known, the system state
can be updated by numerical integration. As with any other set of
simultaneous first-order ordinarly differential equations, a variety
of integration schemes can be used~\cite{numericalrec1992}, and here a
second-order Verlet algorithm is used. The size of the
integration time step varies throughout each simulation according to the
Courant-Friedrichs-Lewy (CFL) condition~\cite{sphysicsguide2010} and
the magnitude of the interactions. Here, minimum and maximum values of
certain physical quantities need to be found among all particles, and
for this, efficient CUDA reduction algorithms provided by NVIDIA are
employed.

\subsection{Single-GPU versus single-CPU approaches}
\label{sec:gpuSpeedup}


Previous work concerning HPC for SPH published prior to the 
appearance of GPUs report results using either small CPU clusters 
or large supercomputers, such as Blue-Gene.  It is generally misleading to compare the performance of a specific number of GPU cores to a CPU core as the architecture and potential performance is quite different.  However, when demonstrating the advantages of GPU computations for engineering applications within industry, the ability to contrast the relative runtimes can be both accessible and useful.  Hence, results 
can be expressed in terms of speedup and efficiency by comparing the number 
of cores versus a single core. Therefore, to analyse the performance of one GPU, 
the speedup in comparison with a single CPU core is also shown to give 
an idea of the order of speedup that is possible when using low-cost
and accessible GPU cards, instead of large cluster machines.


We have observed (see Ref.~\cite{crespo2011}) that the GPU version of 
DualSPHysics runs on the order of 60 times 
faster than its single-threaded CPU version. For this comparison, an NVIDIA GTX480 GPU
and an Intel Core i7 were used. In the CPU code, all of the standard CPU optimizations were applied, 
symmetry in particle-particle interaction was employed, and SIMD instructions 
that allow performing operations on data sets were used whenever possible.
However, if a multi-threaded, OpenMP
approach 
is used on a multi-core CPU instead (4 cores of a CPU Intel Core i7, 
with 8 logical cores using Hyperthreading) the speedup of the GPU over the CPU was 14.
In this way, for the mentioned hardware one can estimate a speedup of approximately $60/nc$
where $nc$ is the number 
of CPU cores used.

\begin{equation*}
\end{equation*}

\section{Multi-GPU implementation}
\label{sec:comptechmultip}

As mentioned in Section~\ref{sec:comptechsingle}, single-GPU
DualSPHysics has proved to be a viable option for accelerated SPH
simulations.
However, for a
  fully-on-GPU approach as presented here, there is a limitation on the
number of particles that can be simulated due to the size of the GPU
memory. For example, a commonly used NVIDIA GPU, the GTX480, with
1.5~GB of memory, can handle up to about 8 million SPH particles, and an
NVIDIA Tesla M2050 can simulate a maximum of about 15 million
particles.  To go beyond the limits\footnotemark imposed by those constraints, 
 a spatial decomposition scheme is introduced
(illustrated in Fig.~\ref{fig:ima_DANIEL0050}) with the use of MPI to
communicate between different GPUs, which is explained in this
section.

\begin{figure}
  \centering
  \includegraphics[height=12.0cm,angle=0]{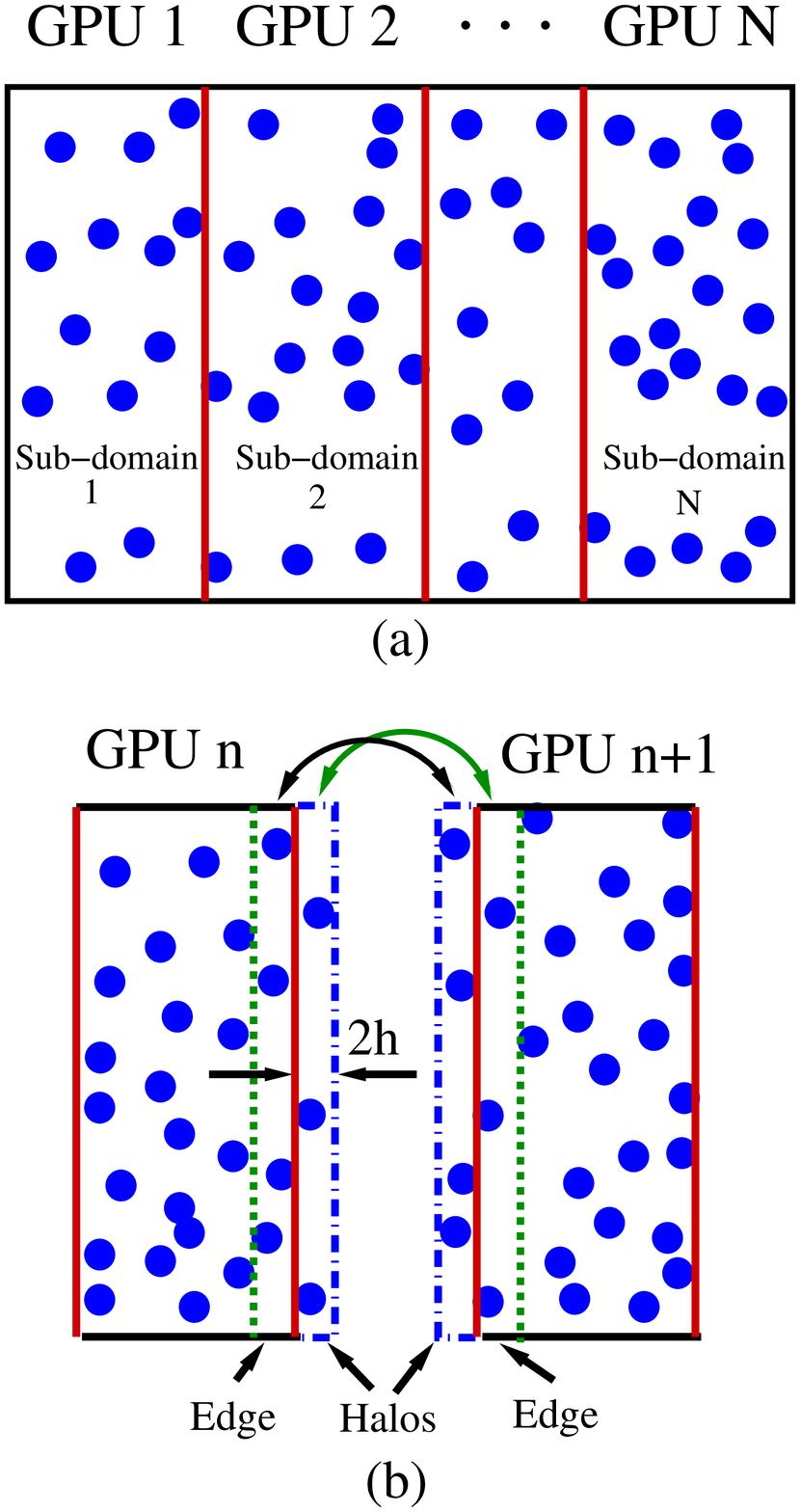}
  \caption{(a) One-dimensional domain decomposition scheme for a
    computational system with $N$ GPUs. The total physical volume is
    divided into $N$ sub-domains, each of which is assigned to a
    different GPU. Data needs to transferred between GPUs when there
    is flow from one sub-domain to another, and also when the halo
    needs to be updated. (b) Data needed to process the dynamics of
    particles within the range of interaction (a distance of twice the
    smoothing length, $2h$, in our case) is stored in the halo of the
    sub-domain.}
  \label{domaindecompDim1both}
\end{figure}

\begin{figure}
  \centering
  \includegraphics[height=5.7cm,angle=0]{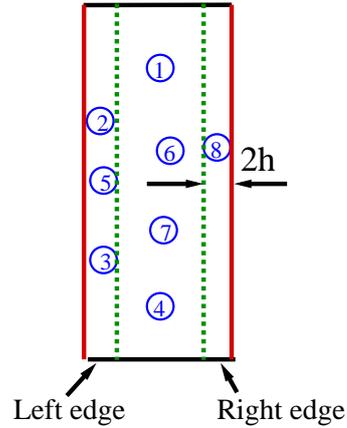}
  \caption{The ``edges'' of a sub-volume are used to construct the halo
    of neighbouring sub-domains. The illustrated sub-domain is assumed
    to have neighbouring sub-domains both to the left and to the
    right, and therefore data arrays containing particle state are
    ordered with the use of radix sort routine, as explained in detail
    in the text.}
  \label{domaindecompDim1radix}
\end{figure}

\footnotetext{An alternative solution to a limited
    size of GPU memory when doing single-GPU simulations is to keep
    the particle information on the host and transfer data to the GPU
    in batches, one after the other, until all particles have been
    processed. It is estimated that this procedure would be quite
    inefficient compared to a multi-GPU approach due to long GPU to
    host data transfer times, and the fact that there would be only
    one GPU processing the data of all particles.}

\subsection{A multi-GPU system}

To understand the computational tasks 
required for a multi-GPU SPH simulation, we begin with a brief
description of our multi-GPU system. Fig.~\ref{fig:gpucpuconnected}
shows a schematic view of one such system, consisting of two computing
nodes, each hosting one CPU and two GPUs. More generally, a multi-GPU
system consists of one or more computing nodes, each hosting one or
more GPUs, in addition to one or more CPUs. Nodes are connected with
each other via a computer networking technology, and the transfer of
data between nodes is executed using a protocol, in this case MPI,
whereas each GPU is connected to its host via a PCI Express bus.
Throughout the rest of this article, the words ``CPU'', ``node'' and
``host'' are used interchangeably.

\subsection{Spatial decomposition}

The main idea behind the implementation of our scheme is to assign
different parts of the physical system to different GPUs---that is, 
 a volume domain decomposition technique is used. The portion
assigned to a GPU is referred to as a ``sub-domain'' throughout the rest of this article.
Eventually, the intention is to use a multi-dimensional domain
decomposition with load balancing. For the development of the
algorithm presented herein, a one-dimensional decomposition is
sufficient.
Fig.~\ref{fig:ima_DANIEL0050} illustrates a domain decomposition scheme in operation for
a three-dimensional dam break simulation with three obstacles. During
and after each integration step, data needs to be transferred between
GPUs due to, firstly, particles migrating from one sub-domain to
another, and secondly, information of particles residing on a neighbour
sub-domain that are close enough to the boundary to influence the
dynamics of particles in the domain of interest, that is, the 'halo
building' process. This requires three main steps: preparation of the
data, as well as GPU-CPU communications, and inter-CPU
communications. Preparation of data is the process by which a GPU
re-arranges the information to be passed to other GPUs
before actually sending it. An example of this is the re-ordering
 of the arrays containing the state
of particles when they step out of their sub-domain, so that the
relevant information is efficiently packaged before it is copied to the
CPU (host) memory.

\begin{figure}
  \centering
  \includegraphics[height=5.5cm,angle=0]{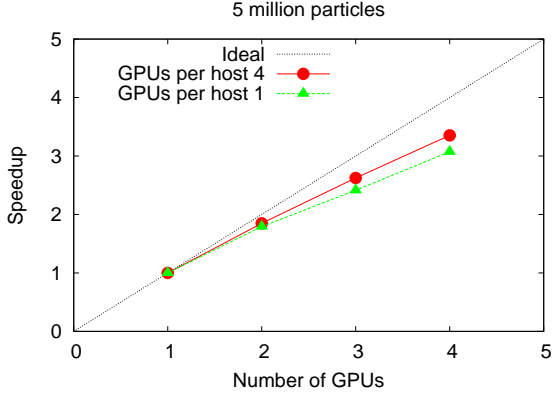}
  \caption{Strong scaling behaviour for the speedup of a dam break
    simulation with 5 million particles. The red curve corresponds to
    simulations run on the system with up to four GPUs, all residing
    on the same node, whereas the green curve shows results using one
    GPU per node.}
  \label{fig:speedup_05m}
\end{figure}

Fig.~\ref{domaindecompDim1both}a illustrates a one-dimensional
domain decomposition scheme for a computational system with $N$
GPUs. The total physical volume is divided accordingly into $N$
sub-domains (boundaries are shown in red in Fig.~\ref{domaindecompDim1both}),
each of which is assigned to a different GPU. 
Fig.~\ref{domaindecompDim1both}b shows two neighbouring
domains, $n$, and $n+1$. Data from the \emph{edge} (the portion of the
sub-domain within a distance $2h$ to the sub-domain boundary) of
domain $n$ is used to construct the \emph{halo} of the neighbouring
domain $n+1$, and vice versa.  The width of the halo and the width of
the edge are the size of the interaction range, in our case, twice the
smoothing length, $2h$.

For the sake of clarity, in the rest of this sub-section we assume a
muti-GPU system with $N$ nodes. Each node is labelled with the number $k$
  (with $k=1,2,\ldots N$) and hosts only one CPU (CPU$_k$) and one GPU
  (GPU$_k$). There is, then, one MPI process, $p_k$, per GPU, and
  $p_k$ controls GPU$_k$, which is assigned to sub-domain $S_k$.
After $p_k$ sets the initial state $Q_k$ (position, velocity, etc.) of
particles in $S_k$ {\it and} its halo $H_k$, the program iterates over
the following steps:
  \begin{itemize}
  \item Determine integration step size, $\Delta t$ 
  \item GPU$_{k}$: update state, $Q_{k}$
  \item GPU$_{k}$: if any particles now have positions corresponding to a different sub-domain $S_j$, with $j\neq k$, then
    \begin{itemize}
    \item GPU$_{k}$: label particles according to sub-domain 
    \item GPU$_{k}$: order arrays according to sub-domain label (using radix sort)
    \item GPU$_{k}$: copy data, $D_{send}$, from GPU$_k$ to CPU$_k$, of particles migrating from one sub-domain to another
    \item CPU$_{k}$: send $D_{send}$    to   $p_j$, with $j\neq k$ 
    \item CPU$_{k}$: receive data, $D_{receive}$, from $p_j$, with $j\neq k$ 
    \item GPU$_{k}$: copy $D_{receive}$ from CPU$_k$ to GPU$_k$
    \item GPU$_{k}$: update number of particles in sub-domain $k$
    \end{itemize}
  \item if saving data, then
    \begin{itemize}
    \item GPU$_{k}$: copy data of all particles in sub-domain from GPU$_k$ to CPU$_k$
    \end{itemize}
  \item GPU$_{k}$: label particles according to edge 
  \item GPU$_{k}$: order arrays according to edge label (using radix sort)
  \item GPU$_{k}$: copy data, $E_{send}$, of edge particles, from GPU$_k$ to CPU$_k$
  \item CPU$_{k}$: send $E_{send}$    to   $p_j$, with $j\neq k$ 
  \item CPU$_{k}$: receive data, $E_{receive}$, from $p_j$, with $j\neq k$ 
  \item GPU$_{k}$: copy $E_{receive}$ from CPU$_k$ to GPU$_k$
  \item GPU$_{k}$: update number of particles in sub-domain $k$
  \item GPU$_{k}$: order arrays according to cubic cell of size $2h$
  \end{itemize}
  Determining the size of $\Delta t$, which is the same for all sub-domains,
  involves two kinds of reduction operations: one on the GPU, where
  each device GPU$_k$ determines the minimum size of $\Delta t_{k}$ suitable for the
  numerical integration for particles in its sub-domain, and then
  another, on the CPUs, whereby the minimum among all $\Delta t_k$'s is found with an MPI
  Allreduce operation.
  
\begin{figure}
  \centering
  \includegraphics[height=5.5cm,angle=0]{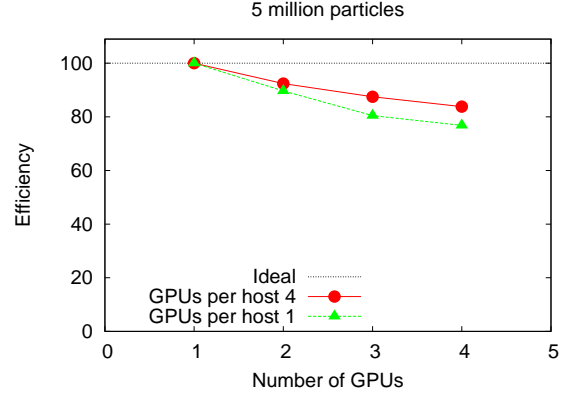}
  \caption{Efficiency for the simulations described in Fig.~\ref{fig:speedup_05m}.}
  \label{fig:efficiency_05m}
\end{figure}

In summary, the multi-GPU program iterates over the next three main steps:

\begin{enumerate}
\item Update particle state: Each GPU updates the state of the
  particles within the sub-domain assigned to it. This involves (i)
  re-ordering data arrays that hold the state of the particles
  to determine efficiently their neighbours (using
  the radix sort algorithm), (ii) computing the inter-particle
  interactions (e.g., the forces), and (iii) the
    actual update of the particle state.
\item Update sub-domains: the number of particles in each sub-domain
  must be updated to reflect particle migration due to fluid flow, and
  the state of migrated particles must be transferred
  accordingly. This is achieved by re-ordering data arrays (using the
  radix sort algorithm) according to the sub-domain in
    which they are located, and then transferring the necessary data
    from the GPU to the CPU\footnotemark, then among CPUs if the GPUs
    reside on different hosts, and last, from the CPU to the new GPU.
\item Update sub-domain halo: The halo holds the data of particles
  that exist on a neighbouring sub-domain but close
  enough to influence the particles of the domain in question (a
  distance twice the smoothing length $h$ in our case, see
  Fig.~\ref{domaindecompDim1both}b). Here again the
  radix sort algorithm is used to re-order data arrays according to the halo to which they
  they may belong.
\end{enumerate}

In its current version, the scheme does not include task overlapping of computation on each GPU with communications among those devices. 
Since the computation of the
particles inside a given sub-domain requires an up-to-date state 
of the particles
within its halo, it is not possible for a given GPU to send data 
to another device before the computation is finished, which prevents the overlap of
computation and communication. 

In regards to overlapping of CPU and GPU computation, 
the highly parallelizable
nature of the SPH algorithm, in conjunction with the high cost of GPU-CPU communications,
 makes fully-on-GPU schemes more efficient, at least in single-GPU computations. This has been 
discussed in Ref.~\cite{ogerSPHERIC2010} (e.g., see Figure 1 of that work). However, 
it is conceivable that in multi-GPU approaches, due to the fact that data needs to be transferred
among GPUs via the CPU\footnotemark, one could perform some of the computations on the CPU, thereby saving
GPU-to-CPU communication time. This idea remains to be explored.

\footnotetext{At the time of writing, May 2011, CUDA
    4.0 Release Candidate has become available, and the production
    version is expected soon. CUDA 4.0 allows direct
    communications among multiple GPUs, and MPI transfers directly
    from the GPU memory without an intermediate copy to system (CPU)
    memory. However, this capability is only available to NVIDIA's
    Tesla GPUs with the Fermi architecture (e.g., Tesla M2050)
    and GPUs must reside on the \emph{same node.}}


\begin{figure}
  \centering
     \includegraphics[height=5.5cm,angle=0]{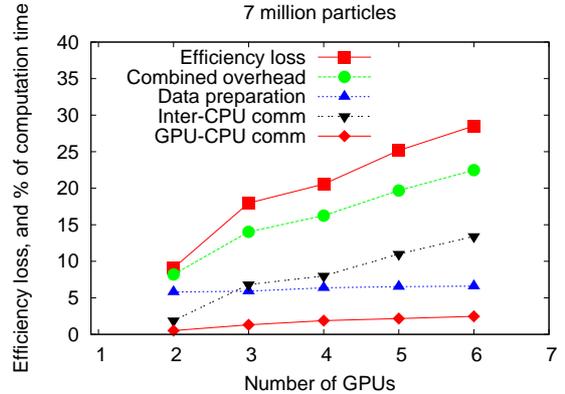}
  \caption{Efficiency loss is compared to the percentage of time
    consumed by various computational tasks in a multi-GPU
    simulation.}
  \label{fig:times}
\end{figure}

\subsection{Using the radix sort algorithm}

To illustrate the use of the radix sort algorithm for the halo
updating procedure, we use Fig.~\ref{domaindecompDim1radix}. It is
assumed that the sub-domain represented in this figure has neighbour
sub-domains both to the left and to the right. Particles labelled 2, 5,
and 3 are within a distance $2h$ from the sub-domain boundary to the
left and therefore form the left ``edge'', whereas particle labelled $8$
is within this distance to the sub-domain boundary to the right, and
forms the right ``edge''. Particles 1, 4, 6, and 7 are further than $2h$
from either boundary and therefore do not belong to any edge. When the
update of the halo of the sub-domain neighbouring to the left is made,
the state of particles on the left edge needs to be passed to the GPU
assigned to it. To achieve this, first a CUDA kernel is launched which
assigns to particles \emph{not} in a halo the label ``0'', and to particles on
the left halo a label ``1'', and to particles in the right halo the
label ``2'':
\begin{eqnarray}
  halo &=& \{\textcolor{red}     {0,\, 1,\, 1,\, 0,\, 1,\, 0,\, 0,\, 2}\},       \nonumber \\
  id  &=&  \{\textcolor{blue}      {1,\, 2,\, 3,\, 4,\, 5,\, 6,\, 7,\, 8}\},       \nonumber 
\end{eqnarray}
where ``halo'' and ``id'' are one-dimensional data arrays of integers,
containing the assigned label, and the particle identification number,
respectively.

Next the radix sort routine is called to obtain a re-ordered array of
the particles identification numbers, according to the ``halo'' label:
\begin{eqnarray}
  halo &=& \{\textcolor{red}     {0,\, 0,\, 0,\, 0,\, 1,\, 1,\, 1,\, 2}\},       \nonumber \\
  id  &=&  \{\textcolor{blue}      {1,\, 4,\, 6,\, 7,\, 2,\, 3,\, 5,\, 8}\}.       \nonumber 
\end{eqnarray}
The new version of the array ID is used to re-order all of the
arrays containing the state of the particles, namely, the one
dimensional array for positions, velocities, etc. Once this reordering
is made, the CUDA function \emph{cudaMemcpy} is used to transfer the
relevant data from the GPU to the CPU, and, if the GPUs reside on
different hosts, then an inter-node communication is made before
uploading the data in question onto the desired GPU.

A similar procedure to the construction of the halo is used for the data
transfers due to particle migration. In this case, particles are
assigned labels depending on the sub-domain to which they have
migrated: ``1'' if they migrated to left, ``2'' if they migrated to the
right, and ``0'' otherwise, and data is transferred accordingly after
reordering the data arrays containing the particle state, similar to the
construction of the halo.

\subsection{Other approaches to multi-GPU programming}

%


There are three main features which make our scheme
different to other multi-GPU approaches for SPH (for example,
\cite{ogerSPHERIC2010}): 
\begin{enumerate}
\item the use a low-level CUDA approach
instead of a high-level, directive-based transformation of CPU to GPU
code; 
\item the full implementation of the computations on the GPU,
instead of porting to the GPU only part of the computations, and
performing the rest on the CPU; and 
\item the fact that the code
development starts from a single-GPU code and progresses towards a
multi-GPU application, instead of going from a multi-CPU to a
multi-GPU program.
\end{enumerate}

The use of a low-level approach, combined with the fact that the 
computation is fully implemented on the GPU [points (1) and (2)], facilitates a higher 
level of control of the hardware, the utilization of the memory 
hierarchies, as well as the use of highly optimized 
CUDA functions. These features are not as readily available in directive-based approaches.
Additionally ---and in regards to multi-GPU programming--- the advantage 
of these two features is
the straightforward availability of both data and
optimized CUDA sub-routines to process it, for instance, during the
construction of the halo and the migration of particles, which are crucial 
steps in the scheme.

The advantage of point (3) is related to the previous 
two points, where the data structures and tools for the multi-GPU scheme were present 
and thoroughly tested on the single-GPU version before
making the extension to multi-GPU. For instance, the routines for sorting
and determining the labels of the first and last particle in a given region of space
were already used
in single-GPU method for finding neighbours of particles, and when implementing the
multi-GPU version they were re-used for determining migrating particles as well as 
the edges and halos in a sub-domain, as explained in detail above.


\section{Hardware}
\label{sec:hardware}

Simulations have been performed on two different systems: one is part
of a dedicated cluster at the University of Manchester, where the GPU
section has six nodes each hosting two NVIDIA Tesla M2050, which
feature the Fermi architecture, and each possess 448
computing cores grouped in 14 streaming
multiprocessors of 32 cores each, clock rate of 1.15 GHz, 3 GB GDDR5 memory, and a high
speed, PCI-Express bus, Gen 2.0 Data Transfer. The hosting nodes are connected via 1 Gigabit per
second Ethernet technology, which, unfortunately, is much slower than
Infiniband, with its current speed of 40 Gigabits per second and
significantly lower latency, to which we currently have no access. 
Each of the hosting nodes also has two six-core AMD
Opteron processors (CPUs), and runs Scientific Linux release 5.5.

The second system used for simulations is a single
node hosting four GPUs at the University of Vigo, 
Spain. The GPUs are NVIDIA GTX480, which, like the M2050, feature a
Fermi architecture and 448 computing cores, but have a faster clock rate of 1.4
GHz, and a smaller memory of 1.5 GB GDDR5. The connection to the host is done
via a PCI-Express bus. Additionally, the hosting node has two Intel
Xeon E5620 2.4 GHz CPUs, and since all four GPUs reside on the same
host, there is no need for a computer networking technology.

\section{Results}
\label{sec:results}

To test our multi-GPU program, simulations of a three-dimensional dam
break with three obstacles, illustrated in
Fig.~\ref{fig:ima_DANIEL0050} were performed. When assessing
parallelization, it is useful to address questions of efficiency and
scalability in order to determine the usefulness of the approach
used. The aim of this section is to answer the following
questions:
\begin{itemize}
  \item Scaling: how do computing times change when the system size
    and the number of computing devices (in our case, the number of
    GPUs) vary?
  \item Where are the bottlenecks? Are they in the GPU to CPU or in
    inter-CPU data transfers? Or are they on the data preparation routines?
  \item How does the overhead resulting from data transfers between
    GPUs (due to particle migration and halo construction) compare to
    the time spent by each GPU processing the motion of particles in
    its assigned sub-domain? 
\end{itemize}

\begin{figure}
  \centering
      \includegraphics[height=5.0cm,angle=0]{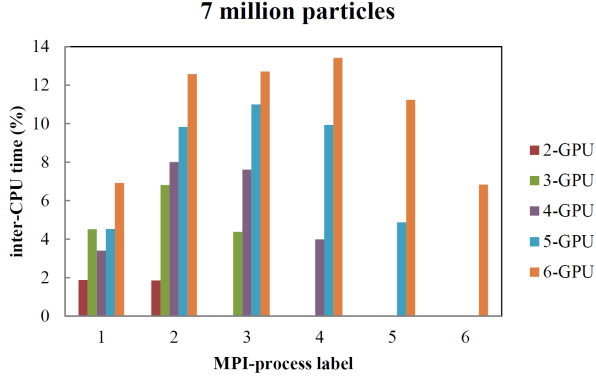}
  \caption{Percentage of time that each MPI process uses for inter-CPU
    communications for simulations using different number of GPUs. The
    horizontal axis corresponds to the label of the MPI
    process. Smaller times correspond to the processes assigned to
    endpoints of the simulation box, as explained in more detail in
    the text.}
  \label{fig:inter_CPU}
\end{figure}

The speedup $s$ of the multi-GPU simulations is defined as a function
of the number of GPUs $N$ in the following way:
\begin{equation}
  s(N)=\frac{T(N_{\textsc{ref}})n_p(N)}{T(N)n_p(N_{\textsc{ref}})}
  \label{eqn:speedupgeneral}
\end{equation}
where $T$ is the simulation time \emph{divided by the number of
  steps}\footnotemark~$N_{steps}$, $N_{\textsc{ref}}$ is the number of
GPUs used as reference (in this article, $N_{\textsc{ref}}=1$), and
$n_p$ is a measure of the system size, which in this article is the
number of particles.  The efficiency $\eta$ is defined as the speedup
divided by the number of GPUs:
\begin{equation}
  \eta(N)=\frac{s(N)}{N}.
  \label{eqn:efficency}
\end{equation}
Two types of scaling behaviour are investigated, \emph{weak} and
\emph{strong}, which are now described.

\footnotetext{The simulation time per step $T$ is used instead of the
  simulation time $T_{sim}$ because $N_{steps}$ varies with $n_p$ for
  simulations of a fixed {\it physical time} $t$, and fixed physical system
  dimensions. The reason for this is that
  $N_{steps}=t\;/\;\overline{\Delta t}$, where $\overline{\Delta t}$
  is the average integration step (average over the whole simulation,
  since an adaptive time step algorithm is used), and
  $\overline{\Delta t}$ is typically proportional to the
  discretization length $\Delta p$ (the ``size'' of the SPH particle),
  which determines the number of particles $n_p\propto 1/\Delta
  p^3$. Therefore, for fixed physical time $t$, it is expected that
  $N_{steps}\propto n_p^{1/3}$. Since the simulation time should be
  proportional to both the number of particles and to the number of
  steps ($T_{sim}\propto N_{steps} n_p$), simulation
  times will follow $T_{sim} \propto n_p^{4/3}$. This behaviour has been observed 
   in our simulations.}

\subsection{Strong scaling}

Strong scaling is studied by increasing the number of GPUs while leaving
the system size (both the number of particles and the physical
dimensions) fixed
$n_p(N)=n_p(N_{\textsc{ref}})$. Substituting this in 
(\ref{eqn:speedupgeneral}) we obtain
\begin{equation}
  s(N)=\frac{T(N_{\textsc{ref}})}{T(N)}.
  \label{eqn:speedupgeneralfixsize}
\end{equation}
In the ideal case the computation time is inversely proportional to
the number of GPUs used, $T(N)=c/N$, where $c$ is the proportionality
constant, giving $s(N)=N/N_{\textsc{ref}}$. If the
reference number of GPUs $N_{\textsc{ref}}=1$, then $s(N)=N$.

\begin{figure}
  \centering
  \includegraphics[height=7.0cm,angle=0]{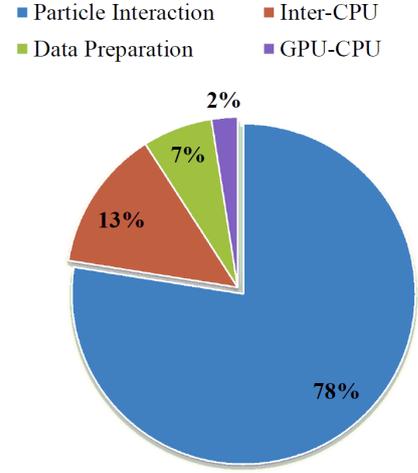}
  \caption{Comparison of the percentage of the computation time spent
    by different tasks carried out during a 4-GPU simulation of a dam
    break, using 7 million particles, where each GPU resides on a
    different computing node.}
  \label{fig:computingtimeschart}
\end{figure}

Fig.~\ref{fig:speedup_05m} shows results for the strong scaling
behaviour of our program for the speedup, for a dam break simulation
of 5 million particles, on the two systems described in
Section~\ref{sec:hardware}. 5 million particles was chosen because
this number is close to the maximum number of particles that can
currently fit into a single NVIDIA GTX480 used for comparisons
here. The red curve corresponds to simulations run on the system with
up to four GPUs, all residing on the same node, whereas the green
curve shows results using one GPU per node. As expected, the speedup
is better in the first case, where there is no need for inter-CPU
(inter-node) communications. 
Fig.~\ref{fig:efficiency_05m} shows
the corresponding results for the efficiency for the system described in Fig.~\ref{fig:speedup_05m}.

In Fig.~\ref{fig:times}, the efficiency loss, defined as $100-\eta$
(where $\eta$ is the efficiency, plotted in
Fig.~\ref{fig:efficiency_05m}), is compared to the percentage of time
consumed by various computational tasks in a multi-GPU simulation. From this, it can be observed that 
(i) a correlation exists between the combined overhead (sum of
inter-CPU, GPU-CPU communication and data preparation times, in green circles)
and the loss of efficiency (in red squares), and (ii) that the overhead
increases with the number of GPUs mostly due to the inter-CPU
communications (black triangles), since the overhead caused by GPU-CPU
communications (red diamonds) and data preparation (blue triangles) increases
much more slowly. Note that, since each MPI process in a given
simulation measures its own communication times, and those times are
in general different from one process to another, here 
the longest time among all processes of a given simulation is used. This will
be further discussed for Fig.~\ref{fig:inter_CPU} below.

In future versions of our program, the
total overhead is expected to be reduced  with (a) the use of pinned memory for faster the
GPU-CPU data transfers, and (b) the utilization of an Infiniband
switch, which should substantially decrease the inter-CPU
communications. Also, for the case of GPUs residing on the same host,
the introduction of CUDA 4.0 should result in further reduction of
overhead.  As mentioned earlier, two-dimensional domain decomposition is envisaged, which will entail a significant
increase of the inter-CPU communication times, especially on systems
that use slow inter-CPU networking technology, such as Gigabit Ethernet, 
instead of Infiniband or faster versions of Ethernet.
The reason for this is that for two-dimensional domain
decomposition, each MPI process will need to send information to as
many as eight other processes when doing the inter-CPU communications
on system with many nodes.

\begin{figure}
  \centering
    \includegraphics[height=5.5cm,angle=0]{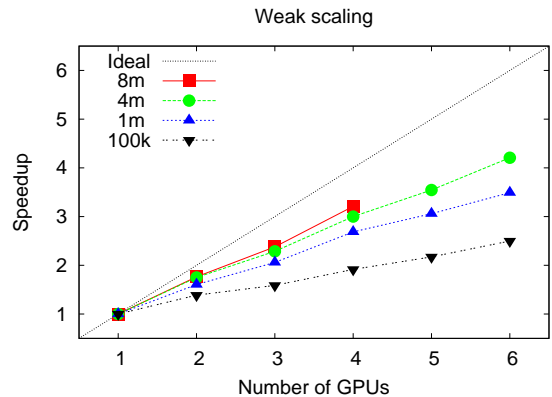}
  \caption{Simulations using a number of particles proportional to the
    number of GPUs. Plots are shown for $n_{sub}$ 
    (number of particles per GPU) equal 100 thousand, 1, 2, 4, and 8
    million. 
}
  \label{fig:speedup_weakscaleNGpuPNode1}
\end{figure}

Figure~\ref{fig:inter_CPU} shows the percentage of time that each MPI
process uses for inter-CPU communications for 2-, 3-, 4-, 5-, and
6-GPU simulations. The horizontal axis corresponds to the label of the MPI
process. Each label is assigned in order of increasing $y$-coordinate
to the 'slices' into which the simulation box has been subdivided
(e.g., Fig.~\ref{fig:ima_DANIEL0050} for the particular case of three
GPUs).  One can see that the time is smaller for the processes
assigned to sub-volumes on the edges of the simulation box than for
the rest, reflecting the fact that processes on the edge need to send
data to one neighbour process only, whereas the rest of them need to
send data to two neighbours instead\footnotemark.

\begin{figure*}
  \centerline{
    \subfigure{\includegraphics[width=6.0in]{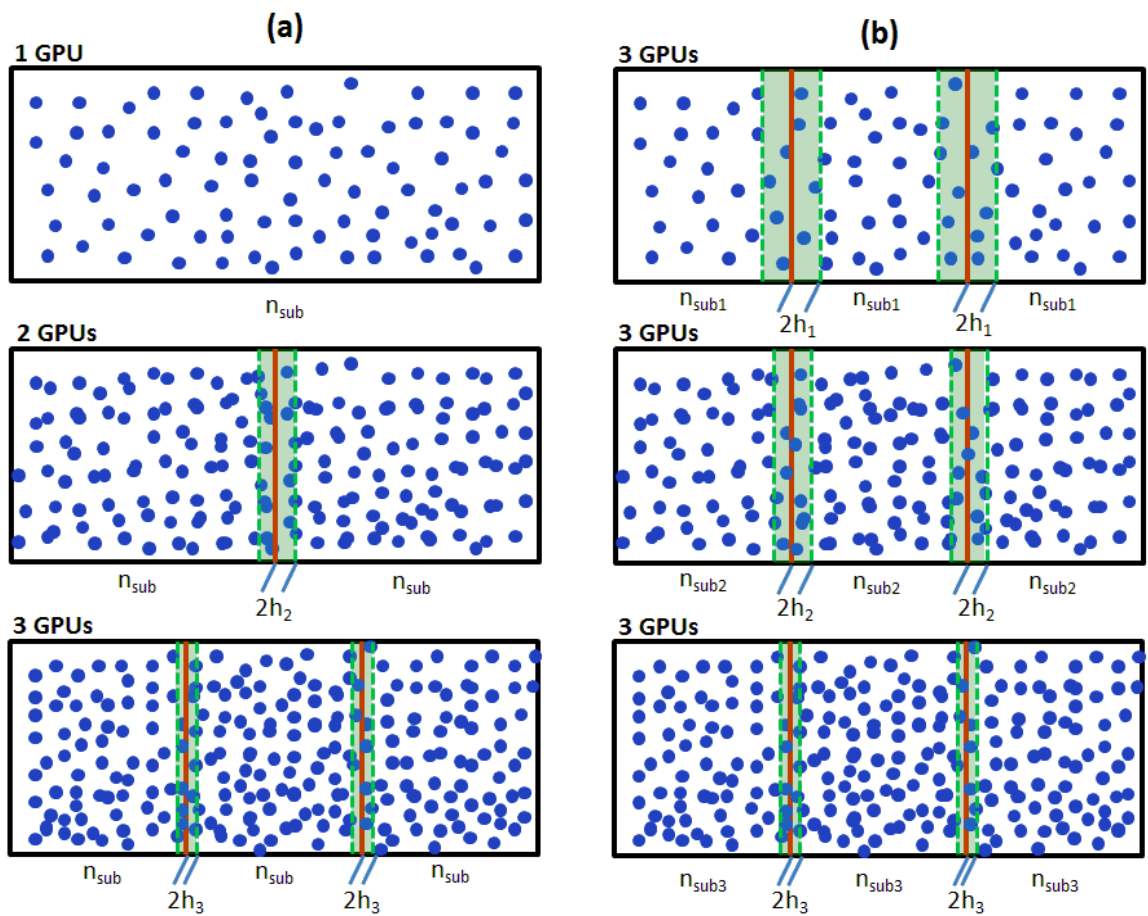}}
  }
  \caption{(a) Domain decomposition scheme for a fixed number of 
particles $n_{sub}$ per GPU,  and an increasing number of GPUs, corresponding to 
weak scaling. 
(b) For a given number of GPUs (three in this example), one can show that 
the ratio of particles in the halo $n_{halo}$ to $n_{sub}$ decreases, leading to the better
scaling that can be seen in 
Fig.~\ref{fig:speedup_weakscaleNGpuPNode1}. See text for details.}
  \label{fig:Figure12_full}
\end{figure*}

\footnotetext{We have observed that CPU data preparation times are
  also shorter on processes assigned to simulation box edges than on
  the rest. GPU data preparation times are, on the other hand, flat
  across all processes of a given simulation, which is also consistent
  with the way in which the GPU prepares data. Data preparation times
  in Fig.~\ref{fig:times}  are the sum of GPU and CPU
  preparation times of the highest time-consuming process.}

In Fig.~\ref{fig:computingtimeschart} some of the most
consuming tasks of a multi-GPU simulation are compared for a dam break with 7
million particles on four GPUs. 
This figure shows that
the time that each GPU
uses to compute the particle dynamics 
(once its halo has been updated) is nearly 80\% of the total simulation time,
dominating the overall computation. However, considerable time
is also spent on inter-CPU communications, and, to a lesser extent,
data preparation, and GPU-CPU data transfers.

\subsection{Weak scaling}

Simulations have also been performed using system sizes proportional
to the number of GPUs : $n_p(N)=Nc$, where
$c$ is the proportionality constant. Substituting this in
(\ref{eqn:speedupgeneral}) we get
\begin{equation}
  s(N)=\frac{T(N_{\textsc{ref}}) N}{T(N) N_{\textsc{ref}}}.
  \label{eqn:speedupgeneralfixsize2}
\end{equation}
In the ideal case the time is the same regardless of the number of GPUs,
$T(N)=T(N_{\textsc{ref}})$ and 
$s(N)=N/N_{\textsc{ref}}$. If the reference number of GPUs $N_{\textsc{ref}}=1$, then
$s(N)=N$.

Fig.~\ref{fig:speedup_weakscaleNGpuPNode1} shows the results of
simulations using a number of particles proportional to the number of
GPUs. Plots are shown for 100 thousand, 1, 2, 4, and 8 million
particles per GPU, with the largest simulation so far being for 32
million particles distributed among four GPUs. As expected, 
 scaling 
improves with a growing number of particles 
per GPU
because the time spent on
tasks that result in an overhead becomes a smaller percentage of the
total computational time, which, as shown in
Fig.~\ref{fig:computingtimeschart}, 
is
dominated by the update of particle states performed by each GPU.

One must note that, for each curve in Fig.~\ref{fig:speedup_weakscaleNGpuPNode1},
the total number of particles in the system is increased {\it by making a finer discretization} of
the continuum, that is, by using smaller particles (smaller smoothing length $h$), instead of  increasing
the physical dimensions of the simulation box (here this is kept fixed).
In this scenario, one can demonstrate that the larger the number $n_{sub}$  of particles per GPU, the smaller
the fraction $n_{halo}/n_{sub}$, which leads to better scaling.

To demonstrate this, consider the ratio
\begin{equation}
  \frac{n_{halo}}{n_{sub}}\sim \frac{L H 2h }{L H W/N}=\frac{N2h}{W},
  \label{eqn:npproptoh}
\end{equation}

where $L$, $H$, and $W$ are the length, height and width of the simulation box, 
$L H 2h$ is the volume of the halo, and $L H W$ is the total domain of the 
simulation box, which, divided by $N$, gives the volume of the sub-domain. 
Since the volume of one particle is proportional to $h^3$, for a fixed fluid 
volume the total number of 
particles in a simulation is given by $n_p=N n_{sub} \sim h^{-3}$. We can therefore write:

\begin{equation}
  \frac{n_{halo}}{n_{sub}}\sim \frac{N^{2/3}}{n_{sub}^{1/3}}.
  \label{eqn:nhaloproptoh2}
\end{equation}

This means that, for a fixed number of GPUs $N$, the number 
of particles in the
halo, relative to the number of particles in the sub-domain, shrinks.
This is illustrated in Fig.~\ref{fig:Figure12_full}. 
In part (a), a fixed number of particles per GPU with an increasing number of
GPUs is shown, corresponding to weak scaling. One can 
observe the reduction of the size of the halo, which is of size 2$h$. Part 
(b) of this figure shows also decreasing halo size as
number of particles is increased while keeping the number of GPUs fixed.

\subsection{Scaling comparison}

It is instructive to compare the scaling results presented in this article with those of other 
multi-GPU and multi-CPU approaches. 
Oger et al~\cite{ogerSPHERIC2010} describe a multi-GPU SPH scheme that 
uses directive-based, partially-ported GPU code. Although that report does not present
the same kind of scaling as we do here,
one can extract some useful information from 
their data. In particular, from their Figure 9, one can obtain an 
approximation to their strong scaling 
behaviour by dividing the time 
their simulation takes on two nodes by the time it takes on four nodes, which 
in the ideal scaling case would yield 2.
In Ref.~\cite{ogerSPHERIC2010} this value is approximately 1.7 (for 155,500 particles) and 1.8 
(for 2,474,000 particles). The same division using our data for five million particles gives 
approximately 1.7, showing consistency between the two schemes. 

A comparison with multi-CPU code is also interesting. For 
example, Maruzewski \emph{et al.} \cite{LMH-ARTICLE-2009-002} report both 
strong and weak scaling
on SPH simulations on as many as 1024 CPU cores, with a number of particles of up 
to 124 million. 
As it is often the case in this 
type of multi-CPU simulations, scaling data (Figures 5, 6, and 7 in 
Ref.~\cite{LMH-ARTICLE-2009-002}) 
shows no significant departure from the ideal value until the number of 
CPUs is relatively large, on the order of eight in this case.

A plausible explanation for the worse scaling behaviour on multi-GPU systems than 
on multi-CPU clusters has been proposed in Ref.~\cite{trott2011} for Molecular 
Dynamics simulations. In essence, the part of the program computing the motion 
of particles is accelerated by the GPU, but the inter-device communications  
are not. Therefore, the time spent on the latter, as a fraction of the total simulation time, 
becomes relatively large more quickly as the number of computing devices is increased.


\begin{equation*}
\end{equation*}

\section{Summary and future work}
\label{sec:summary}

This paper presented a computational methodology to carry out
three-dimensional, massively parallel Smoothed Particle Hydrodynamics
(SPH) simulations across multiple GPUs. This has been achieved by
introducing a spatial domain decomposition scheme into the single-GPU
part of the DualSPHysics code, converting it into a multi-GPU SPH
program.

By using multiple GPUs, on the one hand, simulations can be performed
that could also be performed on a single GPU, but can now be obtained even
faster than on one of these devices alone, leading to speedups of
several hundred in comparison to a single-threaded CPU program.  On
the other hand accelerated simulations with tens of millions of
particles can now be performed, which would be impossible to fit on a
single GPU due to memory constraints in a full-on-GPU
  simulation where all data resides on the GPU.  By
being able to simulate---without the need of large, expensive clusters
of CPUs---this large number of particles at speeds well beyond one
hundred times faster than single CPU programs, our software has the
potential to bypass limitations of system size and simulation times
which have been constraining the applicability of SPH to various
engineering problems.

The methodology features the use of MPI routines and the sorting
algorithm radix sort, for the migration of particles between GPUs as
well as domain 'halo' building. A study of weak and strong scaling
with a slow Ethernet connection shows that inter-CPU communications
are likely to be the bottleneck of our simulations, but considerable
overhead is also produced by data preparation, and, to a lesser
extent, by GPU-CPU data transfers. A possible solution to the overhead
caused by the latter is the use of pinned memory, which so far in our
program remains unused.  The use of Infiniband instead of Ethernet
should reduce the overhead cause by inter-CPU communications, and for
the case of GPUs residing on the same host, the use of the recently
released CUDA 4.0 will be introduced, which should further accelerate
communications. Future work includes also the introduction of a
dynamic load balancing algorithm, a multi-dimensional
  domain decomposition scheme, as well as floating body capabilities.
%

\section*{Acknowledgements}

The authors gratefully acknowledge the support of
  EPSRC EP/H003045/1 and a Research Councils UK (RCUK) fellowship.
  This work was partially supported by Programa de Consolidaci\'on e
  Estruturaci\'on de Unidades de Investigaci\'on (Grupos de Referencia
  Competitiva) funded by European Regional Development Fund
  (FEDER). We would like to also thank Simon Hood of the University of
  Manchester, for his support solving hardware and software issues during the simulations,
  Chris Butler of Super Micro, and Tim Lanfear of NVIDIA for their
  help in setting up our multi-GPU cluster.

\bibliographystyle{elsarticle-num}


\begin{thebibliography}{10}
\expandafter\ifx\csname url\endcsname\relax
  \def\url#1{\texttt{#1}}\fi
\expandafter\ifx\csname urlprefix\endcsname\relax\def\urlprefix{URL }\fi
\expandafter\ifx\csname href\endcsname\relax
  \def\href#1#2{#2} \def\path#1{#1}\fi

\bibitem{LMH-ARTICLE-2009-002}
P.~Maruzewski, D.~Le~Touz\'{e}, G.~Oger, F.~Avellan, {SPH} {H}igh-{P}erformance
  {C}omputing simulations of rigid solids impacting the free-surface of water,
  Journal of {H}ydraulic {R}esearch 48~(Extra Issue (2010)) (2010) 126--134.
\newblock \href {http://dx.doi.org/10.3826/jhr.2009.0011}
  {\path{doi:10.3826/jhr.2009.0011}}.

\bibitem{Ferrari20091203}
A.~Ferrari, M.~Dumbser, E.~F. Toro, A.~Armanini, A new 3d parallel sph scheme
  for free surface flows, Computers and Fluids 38~(6) (2009) 1203 -- 1217.
\newblock \href {http://dx.doi.org/DOI: 10.1016/j.compfluid.2008.11.012}
  {\path{doi:DOI: 10.1016/j.compfluid.2008.11.012}}.

\bibitem{spurzemberczik2009}
R.~Spurzem, P.~Berczik, G.~Marcus, A.~Kugel, G.~Lienhart, I.~Berentzen,
  R.~Männer, R.~Klessen, R.~Banerjee, Accelerating astrophysical particle
  simulations with programmable hardware (FPGA and GPU), Computer Science
  Research and Development 23~(3-4) (2009) 231--239.

\bibitem{amada2004}
T.~Amada, M.~Imura, Y.~Yasumuro, Y.~Manabe, K.~Chihara, Particle-based fluid
  simulation on GPU, in: ACM Workshop on General-Purpose Computing on Graphics
  Processors and SIGGRAPH, 2004.

\bibitem{harada2007sphongpus}
T.~Harada, S.~Koshizuka, Y.~Kawaguchi, Smoothed particle hydrodynamics on
  {GPUs}, Proceedings of Computer Graphics International (2007).

\bibitem{herault2010}
A.~H\'{e}rault, G.~Bilotta, R.~A. Dalrymple, {SPH} on {GPU} with {CUDA},
  Journal of Hydraulic Research 48~(1 supp 1) (2010) 0022--1686.
\newblock \href {http://dx.doi.org/10.1080/00221686.2010.9641247}
  {\path{doi:10.1080/00221686.2010.9641247}}.

\bibitem{website:dualsphysics}
\url{http://dual.sphysics.org}.

\bibitem{crespo2011}
A.~J.~C. Crespo, J.~Dom\'{\i}nguez, A.~Barreiro, M.~G\'{o}mez-Gesteira, B.~D.
  Rogers, GPUs, a new tool of acceleration in cfd: Efficiency and reliability
  on smoothed particle hydrodynamics methods, PLoS ONE\href
  {http://dx.doi.org/doi:10.1371/journal.pone.002068}
  {\path{doi:doi:10.1371/journal.pone.002068}}.

\bibitem{top500}
\url{http://www.top500.org/}.

\bibitem{mcintoshsmith2011}
S.~McIntosh-Smith, T.~Wilson, J.~Crisp, A.~A. Ibarra, R.~B. Sessions,
  Energy-aware metrics for benchmarking heterogeneous systems, SIGMETRICS
  Perform. Eval. Rev. 38 (2011) 88--94.

\bibitem{Vignjevic2010}
J.~C. Campbell, R.~Vignjevic, M.~Patel, S.~Milisavljevic, {Simulation of Water
  Loading On Deformable Structures Using SPH}, {CMES-COMPUTER MODELING IN
  ENGINEERING \& SCIENCES} {49}~({1}) ({2009}) {1--21}.

\bibitem{GomezGesteira2010}
M.~G\'{o}mez-Gesteira, B.~D. Rogers, R.~A. Dalrymple, A.~Crespo,
  State-of-the-art of classical sph for free-surface flows, Journal of
  Hydraulic Research 48 (2010) 6--27.

\bibitem{liuandliubook}
G.~Liu, M.B.Liu, Smoothed Particle Hydrodynamics, a meshfree particle method,
  World Scientific, 2003.

\bibitem{monaghan1982}
J.~J. Monaghan, Why particle methods work, SIAM Journal on scientific and
  statistical computing 3 (1982) 422--433.

\bibitem{acreman2010.1143A}
D.~M. {Acreman}, T.~J. {Harries}, D.~A. {Rundle}, Modelling circumstellar discs
  with three-dimensional radiation hydrodynamics, Monthly Notices of the Royal
  Astronomical Society 403 (2010) 1143--1155.
\newblock \href {http://arxiv.org/abs/0912.2030} {\path{arXiv:0912.2030}}.

\bibitem{crespo2007}
A.~Crespo, M.~G\'{o}mez-Gesteira, R.~A. Dalrymple, Boundary conditions
  generated by dynamic particles in sph methods, Computers, materials and
  continua 5(3) (2007) 173–184.

\bibitem{ogerSPHERIC2010}
G.Oger, E.~Jacquin, M.~Doring, P.-M. Guilcher, R.~Dolbeau, P.-L. Cabelguen,
  L.~Bertaux, D.~L. Touz\'{e}, B.~Alessandrini, Hybrid CPU-GPU acceleration of
  the 3-D parallel code sph-flow, in: Proc 5th International SPHERIC Workshop, Ed. B.D. Rogers,
  2010, pp. 394--400.

\bibitem{radixsortpaper2009}
N.~Satish, M.~Harris, M.~Garland, Designing efficient sorting algorithms for
  manycore GPUs, in: Proceedings of IEEE International Parallel and Distributed
  Processing Symposium 2009, 2009.

\bibitem{numericalrec1992}
W.~H. Press, S.~A. Teukolsky, W.~T. Vetterling, B.~P. Flannery, Numerical
  recipes in C (2nd ed.): the art of scientific computing, Cambridge University
  Press, New York, NY, USA, 1992.

\bibitem{sphysicsguide2010}
M.~G\'{o}mez-Gesteirra, B.~D. Rogers, R.~A. Dalrymple, A.~Crespo,
  M.~Narayanaswamy, SPHysics - development of a free-surface fluid solver- Part 1: Theory and Formulations, Computers and Geosciences, \newblock \href {http://dx.doi.org/10.1016/j.cageo.2012.02.029}
  {\path{doi:10.1016/j.cageo.2012.02.029}} .

\bibitem{trott2011}
C.~R. Trott, L.~Winterfeld, P.~S.~Crozier, General-Purpose molecular 
dynamics simulations on GPU-based clusters (2011). arXiv:1009.4330v2 [cond-mat.mtrl-sci].


\end{thebibliography}

\end{document}